\documentclass[twocolumn,preprintnumbers,superscriptaddress,reprint,apl]{revtex4}

\usepackage{graphicx}% Include figure files
\usepackage{bm}% bold math
\usepackage{amsmath}
\usepackage{amssymb}
\renewcommand{\vec}[1]{\bm{#1}}

\begin{document}

%\preprint{PRL/001-BSD}

\title{X-ray tools for van der Waals epitaxy of bismuth telluride topological insulator films}

\author{Stefan Kycia}
%\email[]{skycia@uoguelph.ca}
\affiliation{Department of Physics, University of Guelph, Guelph, Ontario N1G\,1W2, Canada}
\author{S\'ergio L. Morelh\~ao}
%\email[corresponding author:]{morelhao@if.usp.br}
\address{Department of Physics, University of Guelph, Guelph, Ontario N1G\,1W2, Canada}
\affiliation{Institute of Physics, University of S\~ao Paulo, S\~ao Paulo 05508-090, Brazil}
\author{Samuel Netzke}
%\email[]{snetzke@uoguelph.ca}
\affiliation{Department of Physics, University of Guelph, Guelph, Ontario N1G\,1W2, Canada}
\author{Celso I. Fornari}
%\email[]{celsoifornari@gmail.com}
\affiliation{National Institute for Space Research, S\~ao Jos\'e dos Campos, S\~ao Paulo 12227-010, Brazil}
\author{Paulo H. O. Rappl}
%\email[]{rappl@las.inpe.br}
\affiliation{National Institute for Space Research, S\~ao Jos\'e dos Campos, S\~ao Paulo 12227-010, Brazil}
\author{Eduardo Abramof}
%\email[]{eduardo.abramof@inpe.br}
\affiliation{National Institute for Space Research, S\~ao Jos\'e dos Campos, S\~ao Paulo 12227-010, Brazil}

\date{\today}

\begin{abstract}
Potential applications in spintronics and quantum computing have motivated much recent research in epitaxial films of bismuth telluride. This system is also an example of van der Waals (vdW) epitaxy where the interface coherence between film and substrate is based on vdW bonds instead of strong ionic or covalent bonds. Effects of lattice mismatch on electrical properties and film structure are more difficult to control due to the weakness of the vdW forces. Here we present a general x-ray diffraction method to investigate in-plane atomic displacements and lateral lattice coherence length in vdW epitaxy. The method is demonstrated in a series of films grown at different temperatures and pressures of additional tellurium sources, revealing strong intercorrelations between the lateral features as well as with the n/p-types of free charge carries.
\end{abstract}

%\pacs{Valid PACS appear here}

\maketitle

Van der Waals (vdW) forces play a fundamental role in the emerging field of materials by design \cite{fio14,bhi15,nov16}. Weak vdW interactions drawing together atomic layers allow mechanical exfoliation of layered materials as well as the combination of atomic layers into new heterostructures without the need for lattice matching. In either cases, 2D systems exhibiting unusual physical properties can be created \cite{goy10,gei13,son18}. Epitaxial growth of vdW heterostructures, while providing a high degree of flexibility to combine materials with different crystal lattices \cite{kom92,yyl10,kru11,liu11,gha17,lit17,tan17}, suffer from the absence of strong interlayer forces to dictate the structure's lateral order. As a consequence, surface reconstruction, charge transfers, and built-in electric fields are among a few problems difficult to control in such heterostructures \cite{guo15}. In this sense, it is desirable to have a proper tool to directly probe the lateral structure of 2D materials as a function of growth parameters. For instance, a reliable x-ray diffraction method to access in-plane atomic displacement (disorder), lateral size of crystal domains, and lateral lattice mismatch in epitaxial systems of relevant materials such as bismuth telluride.

Bismuth chalcogenides Bi$_2$X$_3$ (X = Se,Te) are among the most complex materials extensively investigated due to both thermoelectric and topological insulator properties \cite{zha09,che09,hsi09b,has10,and13,xu16,gjs08}. As thermoelectric materials, low lattice thermal conductivity with tunable n- and p-type charge carrier concentration have ensured widely used thermoelectric systems with great figure of merit at temperatures up to $200^\circ$C \cite{gjs08,her14,yuk15,wei16,cho18}. As topological insulators, high-mobility spin polarized surface currents can be produced without external magnetic fields, providing a basic platform for novel physics and devices \cite{hsi09a,zha10,sei13,lhe13,yn16,esc17,kou18}. Intrinsic conduction exclusively through surface states in high quality thin films \cite{wan11,lee12,hoe14,cf16b} have justified the intense research on epitaxy of topological insulators that has take place in the last few years \cite{guo15,kam15,tpg16,bon17,kri17,wan18,ste14,cf16a,gs18}.

Epitaxy of Bi$_2$X$_3$ has an additional complicating factor beyond those expected for typical vdW epitaxy. Rather than periodic structures, layered films can be formed by random stacking sequences of two stable building blocks, X:Bi:X:Bi:X quintuple layers (QLs) and Bi:Bi bilayers (BLs) \cite{ste14,cf16a,gs18,sm17,sm18,lin05}. Standard x-ray diffraction of film reflections along the growth direction have been able to detect the mean number of BLs and to show that the BLs are randomly distributed over many domains in a single film \cite{gs18,sm17,sm18}. This probabilistic occurrence of BLs can create new lateral length scales in terms of atomic disorder and domains sizes, demanding suitable methods to quantify these lateral features. Moreover, structural defects determine the type and density of charge carriers in the bulk \cite{kim15,cho18}. In the films, vacancies and antisite defects have been assigned as responsible for the gradual defect activation from the nominal p-type to n-type in binary end compound without any alloying \cite{bae14}. However, experimental data have shown no direct correlation between defect activation and charge carrier type \cite{yuk15,bae14,cf18}, implying in more complex structure of defects in epitaxial films than can be seen by either local probes or x-ray diffraction of symmetric reflections \cite{ste14,kri17,sm17}.

Currently, there are a large number of x-ray diffraction techniques able to probe surface and interface defects in thin films \cite{sm91,sm93a,sm93b,lha98,mah99,sm98,sm02b,men09,sm07,men10,edp17,ehs17}, superlattices \cite{sm02a,sm02d,sm03}, laterally patterned epitaxial systems \cite{pie04,jzd16}, and even strain and composition of self-assembled nanostructures \cite{fre09,mal11,rof13}. In vdW epitaxy where lateral lattice mismatches do not necessary compromise performance of the final devices, a precise and easy-to-use technique on large batches of samples for optimizing growth conditions is still needed. In this work, an in-house multi-axis single crystal diffractometer is used to access diffraction vectors with different in-plane components of bismuth telluride films grown on BaF$_2$ (111) substrates. A well defined procedure is developed to assure the necessary accuracy to  establish correlations between growth parameters and lateral features of the film structure. It reveals a complex interplay of these lateral features, their impact on crystalline quality and electrical properties of the films. Ultimately, the lateral structure is a key factor to be taken into account when synthesizing vdW epitaxy films.

In single crystals, atomic displacement values are determined by measuring the diffraction power $P_{\rm hkl}=\int I(\theta){\rm d}\theta =I_e|F_{\rm hkl}|^2 N \lambda^3/\sin(2\theta_{\rm hkl})\, V_{\rm cel}$ of different hkl reflections \cite{sm16}; it is also known as integrated intensity of the diffraction curve $I(\theta)$ as a function of the rocking curve angle $\theta$. To apply similar procedure in thin crystalline films, this general expression has to be properly written in terms of the three parameters that are varying from one reflection to another: the scattering angle $2\theta_{\rm hkl}$, the structure factor $F_{\rm hkl}$, and the number $N$ of unit cells within the diffracting volume $NV_{\rm cell}$ for x-ray of wavelength $\lambda$. The scattering intensity by a single electron, $I_e$, also depends on $2\theta_{\rm hkl}$ through the polarization factor $p$ since $I_e\,\propto\,p$. For thin films of uniform thickness and negligible absorption, $N$ is proportional to the beam footprint $A$, leading to
\begin{equation}\label{eq:Phkl}
    P_{\rm hkl}=K\,p\,A\,|F_{\rm hkl}|^2 /\sin(2\theta_{\rm hkl})
\end{equation}
where $K$ is a constant for each sample.

\begin{figure}
  \includegraphics[width=3.4in]{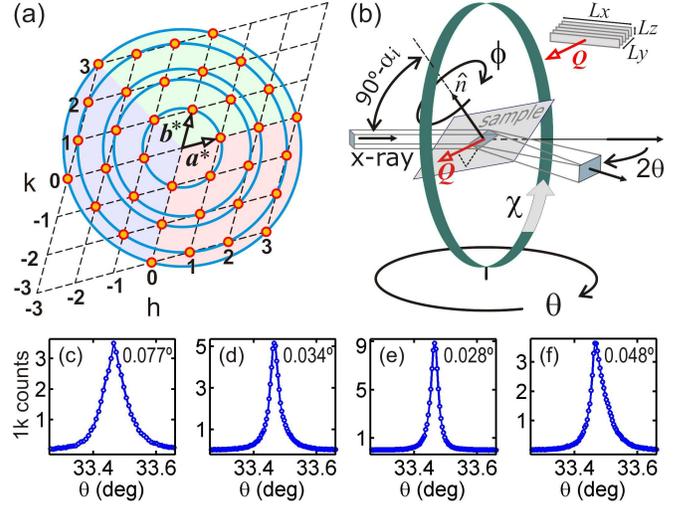}\\
  \caption{(a) In-plane components $\vec{Q}_{y}={\rm h}\vec{a}^*+{\rm k}\vec{b}^*$ of diffraction vectors in relaxed Bi$_2$Te$_3$ (001) films accessible in the single crystal diffractometer with CuK$\alpha_1$ radiation, $Q_{y} =$ 1.656\,\AA$^{-1}$, 2.868\,\AA$^{-1}$, 3.311\,\AA$^{-1}$, 4.381\,\AA$^{-1}$, and 4.967\,\AA$^{-1}$ (blue circles). (b) X-ray footprint at sample surface varying with both $\theta$ and $\chi$ goniometer angles, $\alpha_{\rm i}=\arcsin[\sin(\theta)\sin(\chi)]$ is the incidence angle.  Normal direction $\hat{\vec{n}}$ to the substrate (111) planes is collinear with the rotation axis $\phi$. Diffraction vector $\vec{Q}$ and the scattering angle $2\theta$ are in the horizontal diffraction plane. Inset: definition of crystal grain dimensions, $L_y$ along $\vec{Q}_{y}$ and $L_z$ along $\hat{\vec{n}}$. (c-f) Diffracted intensity curves of reflection $\bar{2}\bar{1}.5$ as a function of the vertical rotation axis $\theta$ in the samples (c) S15p, (d) S17n, (e) S19n, and (f) S27p. Peak widths at half maximum (fwhm) are displayed in each plot.}\label{fig:1}
\end{figure}

When all atoms in the unit cell have similar displacement parameters, the structure factor is simplified to $F_{\rm hkl} = {\rm exp}(-Q_{y}^2 U_{y}^2/2-Q_{z}^2 U_{z}^2/2) \sum_a f_a {\rm exp}(i\vec{Q}\cdot\vec{r}_a)$
where the diffraction vector $\vec{Q} = {\rm h}\vec{a}^* + {\rm k}\vec{b}^* + {\rm l}\vec{c}^*$ has been splitted into two components: $\vec{Q}_{y} = {\rm h}\vec{a}^* + {\rm k}\vec{b}^*$ in the plane of the film, Fig.~\ref{fig:1}(a), and $\vec{Q}_z = {\rm l}\vec{c}^*$ along the growth direction. $U_y$ and $U_z$ stand for in-plane and out-of-plane root mean square atomic displacements, respectively. $f_a$ is the atomic scattering factor of the $a$-th atom at position $\vec{r}_a$ in the unit cell. Within all accessible reflections, there will be only six suitable for measuring atomic displacement in the films, as detailed later in this work.

In epitaxial films composed of many crystal domains with no mosaicity, the line profiles of the diffraction curves, $I(\theta)$, are defined by the distribution of domain sizes. For very asymmetric reflections in non-coplanar rocking curves---sample surface normal direction out of the diffraction plane and close to the rotation $\theta$ axis as in Fig.~\ref{fig:1}(b)---, the most relevant dimension determining the diffraction peak widths is $L_y$ along the in-plane projection $\vec{Q}_y$ of the diffraction vector, inset of Fig.~\ref{fig:1}(b). Domain size effects in the diffraction peaks are accounted for as $I(\theta)\propto\iint|W(\Delta\vec{Q})|^2{\rm d}S$ where ${\rm d}S$ stands for area elements over the Ewald sphere surface at a given angle $\theta$ of the rocking curve and
\begin{equation}\label{eq:WQ}
|W(\Delta\vec{Q})|^2=\prod_{\alpha=x,y,z} \frac{\sin^2(\Delta Q_\alpha L_\alpha/2)}{(\Delta Q_\alpha L_\alpha/2)^2}
\end{equation}
is the normalized volume of each reciprocal lattice node determined by the Fourier transform of the mean domain sizes $L_x$, $L_y$, and $L_z$ \cite{sm16}.

Bismuth telluride films have been grown on BaF$_2$ (111) substrates using a Riber 32P MBE system \cite{cf16a,cf16b}. Besides a nominal Bi$_2$Te$_3$ effusion cell, there are two additional sources of Te. The ratio $\Phi$ between beam equivalent pressures of Te and Bi$_2$Te$_3$ sources can be adjusted to compensate the loss of tellurium during growth. The samples analyzed here are described in Table~\ref{tab:1} with respect to the ratio $\Phi$, substrate temperature $T_{\rm sub}$, and n- or p-type of free charge carrier as obtained by Hall effect.  All films where grown for 2 hours at a constant rate of 0.22\,\AA/s, resulting in thicknesses around 160\,nm as determined either by x-ray reflectometry \cite{cf16b,sm02b} (see supplementary material) or cross-section scanning electron microscopy. The films grow in the trigonal crystal system, space group $R\bar{3}m$, with the (001) planes stacked along the growth direction \cite{wan11,lee12,hoe14,kam15,tpg16,bon17,kri17,wan18,ste14,cf16a}. The in-plane orientation of the films are such that the $\bar{1}0.20$ film reflection falls close to the 331 substrate reflection in reciprocal space \cite{cf16b}, or in terms of real space in-plane directions, $[110]{\rm Bi}_2{\rm Te}_3(001)\,||\,[0\bar{1}1]{\rm BiF}_2$(111) \cite{sm18}.

\begin{table}
\caption{Sample labels, ratio $\Phi$ of beam equivalent pressure between Te and Bi$_2$Te$_3$ sources, substrate temperature ($T_{\rm sub}$), film thickness ($t_f$), carrier density (c.d.), in-plane atomic displacement ($U_y$), mean lateral domain size ($L_y$) and lateral lattice mismatch ($\Delta a/a$).}\label{tab:1}
\scriptsize{
\begin{tabular}{cccccccc}
  \hline\hline
   & & $T_{\rm sub}$ & $t_f$ & c.d. & $U_y$ & $L_y$ & $\Delta a /a$ \\
   Sample & $\Phi$ & ($^\circ$C) & (nm) & ($10^{25}/{\rm m}^3$) & (pm) & (nm) & ($10^{-4}$) \\
  \hline
S15p & 1 & 250 & 165(2) & $+9(2)$ & $16.6(0.1)$ & $54(8)$ & $-6.8(1.0)$ \\
S17n & 1 & 270 & 154(5) & $-4(1)$ & $16.0(0.4)$ & $150(13)$ & $-1.1(0.4)$ \\
S19n & 1 & 290 & 157(10) & $-40(6)$ & $15.8(0.2)$ & $165(18)$ & $+0.1(0.3)$ \\
S27p & 2 & 270 & 160(10) & $+6(1)$ & $15.9(0.3)$ & $79(10)$ & $-5.6(0.5)$ \\
  \hline
  \hline
\end{tabular}
}
\end{table}

X-ray data acquisition was carried out by a Huber four-circle diffractometer sourced by a fine focus copper rotating anode configured with a double collimating multilayer optic followed by a double bounce Ge 220 channel cut monochromator. Bandwidth is 1.2\,eV for CuK$\alpha_1$ ($\lambda=1.540562$\,\AA). Dead time of the sodium-iodide scintillation detector is $\tau = 2.10\,\mu$s; counting rate values are then given by $I = I^\prime \exp(\tau I^\prime)$ for the detector readout values $I^\prime$. Adjustment arcs in the goniometric head were used to orient the diffraction vector of the 222 BaF$_2$ reflection with the $\phi$ rotation axis of the diffractometer within an accuracy better than 0.01$^\circ$. Axial (vertical) divergence is about three times the divergence of 0.005$^\circ$ in the horizontal diffraction plane. Beam cross-section was trimmed down to $0.4\!\times\!0.4$\,mm$^2$ and the detector slits were open wide to accept diffracted x-rays from the full size of the beam footprint at the sample surface. All samples have surface areas larger than $10\!\times\!10$\,mm$^2$.

After the confocal mirrors of the multilayer optics, the beam is still in an unpolarized state before reaching the monochromator. The 220 Ge planes are in the vertical position, implying that after two bounces the horizontal component of the electric field of the incident x-ray beam is reduced by a factor of $\cos^2(2\theta^{\rm Ge}_{220})$ \cite{sm16,sm01,sm02c,sm11}. The horizontal component is further reduced after diffracting in the film, leading to a final polarization factor $p(2\theta_{\rm hkl}) = \left [ 1+\cos^4(2\theta^{\rm Ge}_{220})\cos^2(2\theta_{\rm hkl})\right ]/2$ in the intensity of each hkl reflection of Bragg angle $\theta_{\rm hkl}$.

Diffraction vectors $\vec{Q}$ of asymmetric reflections can be placed in the horizontal diffraction plane by using the $\chi$ and $\phi$ angles, Fig.~\ref{fig:1}(b). The x-ray footprint at the film increases as $\theta$ and $\chi$ differ from 90$^\circ$, since the sample surface plane is vertical for $\chi=90^\circ$ when the surface normal direction $\hat{\vec{n}}$ is set collinear with the rotation axis $\phi$, i.e. when $\hat{\vec{n}} = -[\sin(\theta)\sin(\chi),\,\cos(\theta)\sin(\chi),\,-\cos(\chi)]$. Variation of the footprint in Eq.~(\ref{eq:Phkl}) can be taken into account by using $A = 1/\sin(\alpha_{\rm i})$ where $\alpha_{\rm i}=\arcsin[\sin(\theta)\sin(\chi)]$ is the angle of incidence at the film.

\begin{figure}
  \includegraphics[width=3.4in]{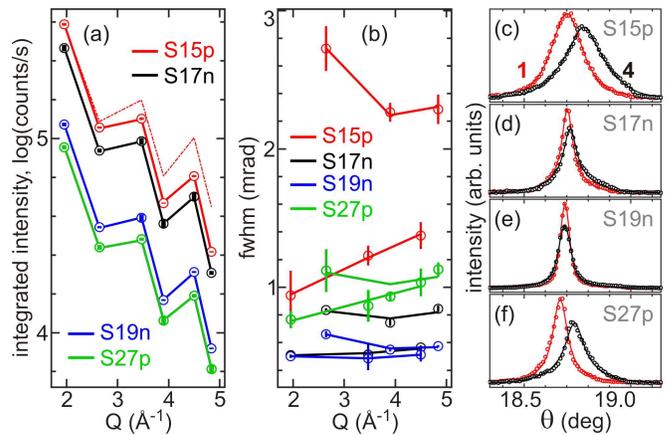}\\
  \caption{(a) Experimental (circles with error bars) and calculated (solid lines) integrated intensities of the asymmetric reflections listed in Table~\ref{tab:2}. For the sake of clarity, two data sets (samples S19n and S27p) are shifted downwards by 0.5 in the log scale. Calculated values (dashed line) without atomic disorder, $U_y=U_z=0$, are also presented. (b) Experimental (circles with error bars) and calculated (solid lines) peak widths at half maximum of the asymmetric reflections. (c-f) Hybrid peaks 1 and 4 used to measure lateral lattice mismatch in the samples \cite{sm18}.}\label{fig:iintfwhm}
\end{figure}

Diffracted intensity curves for one asymmetric reflection as obtained in each sample are show in Figs.~\ref{fig:1}(c)$-$\ref{fig:1}(f). Integrated intensity values (curve area normalized by the counting time) in Fig.~\ref{fig:iintfwhm}(a) correspond to mean values regarding three equivalent reflections that are set apart by 120$^\circ$ rotation in azimuth ($\phi$ axis), as indicated by color-shaded areas in Fig.~\ref{fig:1}(a); for instance, reflections $\bar{1}0.5$, $1\bar{1}.5$, and $01.5$ (or $\bar{1}015$, $1\bar{1}05$, and $01\bar{1}5$ when using hexagonal reflection indexes). Error bars are standard deviations of these three values. A total of 18 reflections were measured on each sample. Theoretical values were fit to the experimental ones by adjusting three parameters: $K$, $U_{y}$, and $U_z$ in Eq.~(\ref{eq:Phkl}). Table~\ref{tab:2} summarizes the goniometer angles and other input values used in the data fitting, which has been carried out by a simulated annealing (SA) algorithm \cite{kir83,kab17}.

\begin{table}
\caption{Input values for fitting diffraction data; used film lattice parameters $a=4.382$\AA\, and $c=30.497$\AA\, \cite{ste14}. Structure factors $F_{hkl}$ were calculated for the Bi$_2$Te$_3$ crystal structure with resonant amplitudes and null Debye-Waller factors. ${\rm h}_s{\rm k}_s{\rm l}_s$ stand for film reflection indexes in the substrate reciprocal lattice \cite{sm18}.}\label{tab:2}
\scriptsize{
\begin{tabular}{cccccccccc}
  \hline\hline
  hkl & $Q$\,(\AA$^{-1}$) & $2\theta_{\rm hkl}$\,($^\circ$) & $\chi$\,($^\circ$) & $\phi$\,($^\circ$) & $p$ & $|F_{\rm hkl}|$ & h$_s$ & k$_s$ & l$_s$\\
  \hline
 $\bar{1}0.5$ & 1.950 & 27.662 & 31.889 & 60.0 & 0.596 & 815.0 & 1.25 & 1.25 & -0.75 \\
 $0\bar{1}.10$ & 2.643 & 37.817 & 51.214 & 120.0 & 0.576 & 725.5 & 2.51 & 0.51 & 0.51\\
 $0\bar{2}.5$ & 3.468 & 50.320 & 17.280 & 120.0 & 0.550 & 695.5 & 3.25 & -0.75 & -0.75\\
 $\bar{2}0.10$ & 3.900 & 57.127 & 31.889 & 60.0 & 0.536 & 637.6 & 2.51 & 2.51 & -1.49\\
 $\bar{2}\bar{1}.5$ & 4.500 & 66.965 & 13.233 & 79.1 & 0.519 & 626.1 & 1.25 & 3.25 & -2.75\\
 $1\bar{3}.10$ & 4.841 & 72.808 & 25.189 & 139.1 & 0.511 & 579.7 & 4.51 & 0.51 & -1.49 \\
  \hline
  \hline
\end{tabular}
}
\end{table}

The measured asymmetric reflections were chosen within the set of accessible ones in Fig.~\ref{fig:1}(a), since they are free of extra intensity contributions from nearby substrate reflections as well as from twinning domains that are often observed in such films \cite{ste14,cf16a}. In reciprocal space, reflections with l = 5 and l = 10 are located at about 1/2 and 1/4 of the distance between adjacent substrate reflections along the surface truncation rods (last three columns in Table~\ref{tab:2}). Intensity contributions from the substrate at these locations are smaller than a factor of $10^{-7}$ regarding the maximum of the nearest substrate reflection \cite{sm17}. Moreover, the beam footprint remains constant within each subset of reflections, i.e. footprint factor $A$ varies only with the reflection index l. For hk.5 reflections $A=7.918$ ($\alpha_{\rm i}=7.255^\circ$), while for hk.10 reflections $A=3.959$ ($\alpha_{\rm i}=14.630^\circ$). It improves the reliability in determining $U_{y}$ since each subset is composed of diffraction vectors with three very distinct in-plane projections, $Q_{y}^2=2.74\,\textrm{\AA}^{-2}$, 11.0\,\AA$^{-2}$, and 19.2\,\AA$^{-2}$, Fig.~\ref{fig:1}(a), diffracting at fixed angle of incidence.

\begin{figure}
  \includegraphics[width=3.4in]{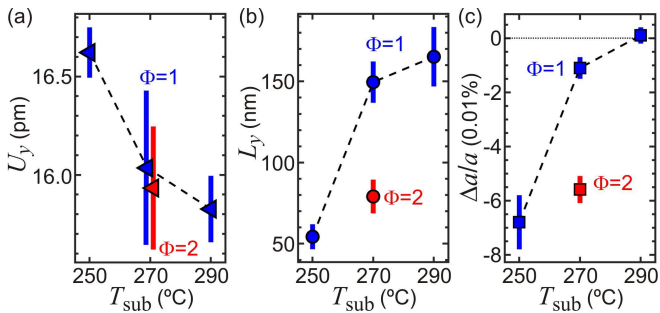}\\
  \caption{Lateral structure of Bi$_2$Te$_3$ films as a function of  substrate temperature $T_{\rm sub}$ during growth and ratio $\Phi$ between beam equivalent pressures of Te and Bi$_2$Te$_3$ sources. (a) In-plane rms atomic displacement, $U_{y}$. (b) Mean-lateral dimension $L_y$ of crystal domains. (c) Lateral lattice mismatch, $\Delta a/a$, displaying direct correlation with the mean-lateral dimension in (b) where $\Delta L_y/L_y = \Delta a/a$. }\label{fig:latparam}
\end{figure}

Values of lateral disorder are presented in Table~\ref{tab:1} (6th column). The uncertainty in $U_{y}$ for each sample was estimated by repeating hundreds of times the data fitting for integrated intensity values randomly distributed within their error bars in Fig.~\ref{fig:iintfwhm}(a). The obtained values of $U_{y}$ are in perfect agreement with the 16.4(0.6)\,pm value (=$\sqrt{U_{11}}=\sqrt{U_{22}}$\,, in ref.~\onlinecite{man14}) for the Te layers around the vdW gap in bulk Bi$_2$Te$_3$ at 300\,K. In films, the limited number of suitable reflections provide only an effective value for all atomic layers, although with enough accuracy to resolve a trend with the growth temperature, e.g. Fig.~\ref{fig:latparam}(a). On the other hand, the obtained out-of-plane disorders of $U_{z}=16(2)$\,pm in all films have poor accuracy due to the small variation of $Q_z$ within the set of measured reflections, $Q_z^2=1.06\,\textrm{\AA}^{-2}$ (l=5) and $4.24\,\textrm{\AA}^{-2}$ (l=10), hindering any observation of possible trends as a function of the growth parameters.

Diffraction peak widths of all reflections are presented in Fig.~\ref{fig:iintfwhm}(b). Error bars are standard deviations from equivalent reflections measured at different azimuths. In non-dispersive diffraction geometry between monochromator and sample Bragg planes \cite{bow98,pie04}, the maximum instrumental broadening is for $0\bar{1}.5$ reflections, yet is no larger than 0.2\,mrad (0.011$^\circ$). Since the experimental widths are much larger than this value, they were directly used to extract mean sizes of diffracting domains through Eq.~(\ref{eq:WQ}). For each subset of reflections, with l=5 or l=10, the $L_x$, $L_y$, and $L_z$ dimensions, were adjusted by the SA algorithm to reproduce the observed widths, as shown by solid lines in Fig.~\ref{fig:iintfwhm}(b). In all fittings, $L_x$ spreads over a range of large values above a few microns, implying in very narrow reciprocal nodes along this direction (perpendicular to $\vec{Q}$ in the horizontal diffraction plane). $L_z$ varies from one subset of reflections to the other, as well as from sample to sample, but without enough accuracy to draw clear trends with growth parameters, see supplementary material for details. $L_y$ values are the most reliable ones according to this fitting procedure, and consistent (within error bars) with both subsets of reflections. The obtained $L_y$ values are presented in Table~\ref{tab:1} (7th column), and plotted as a function of the growth parameters in Fig.~\ref{fig:latparam}(b).

Effects of film/substrate lattice coherence on size of crystal domains in the films can be inferred by accessing the lateral lattice mismatch $\Delta a/a=(a_f-a_s)/a_s$ where $a_s=a_{{\rm BaF}_2}/\sqrt{2}=4.3841$\,\AA\, and $a_f$ is the film in-plane lattice parameter. For relaxed films, the expected value is $a_f=4.382$\,\AA\, \cite{ste14}, which is about 0.05\,\% smaller than the value for bulk material at room temperature \cite{man14}. Here, accurate values of $\Delta a/a$ were obtained by measuring hybrid reflections in this epitaxial system \cite{sm18}. For the hybrid peaks 1 and 4 in Fig.~\ref{fig:iintfwhm}(c), their angular splitting is proportional to $\Delta\theta=-2.035\Delta a/a$. This relationship leads to the $\Delta a/a$ values in Table~\ref{tab:1} (8th column), also shown in Fig.~\ref{fig:latparam}(c) as a function of growth parameters.

Small lattice distortions around point defects such as vacancies and antisites tend to increase the rms atomic displacements. Reduction of atomic disorder in the films with higher growth temperature, as seen in Fig.~\ref{fig:latparam}(a), is consistent with reduction of point defects. Within the resolution of our measurements, atomic disorder in the films depends only on the growth temperature. Films grown at the same temperature show identical values of atomic disorder, although presenting different type of charge carries. This result suggests that point defects are not directly related to the type of charge carries.

Formation of Bi bilayers (BLs) requires vacancies of Te due to desorption as well as enough mobility of Bi. By raising the growth temperature, Bi mobility increases, Te vacancies give rise to BLs, and atomic disorder becomes smaller. BLs in the vdW gap change the in-plane lattice parameter and prevent film relaxation due to weak vdW interactions. Higher pressure of additional tellurium can compensate desorption, avoiding formation of BLs, and allowing relaxation of the film. Strained films (small or null mismatch) have n-type of charge carrier, suggesting the flip between p- and n-types is driven by the presence of metallic bismuth BLs.

Density of defects limiting the film lateral coherence length, i.e. the domains lateral dimensions, are dictated by the lateral lattice mismatch. Exactly as the density of misfit dislocations in semiconductor epitaxy. Therefore, from the perspective of film crystalline quality, lattice matching is also a relevant issue in van der Waals epitaxy.

\begin{acknowledgments}
The authors acknowledge the financial support from Brazilian agencies CAPES (Grant No. 88881.119076/2016-01) and FAPESP (Grant No. 2016/22366-5 and 2018/00303-7), as well as from Natural Sciences and Engineering Research Council of Canada (NSERC)
\end{acknowledgments}

\clearpage
\bibliography{lateraldisorderBi2Te3}

\end{document}

%\preprint{PRL/001-BSD}

\title{Lateral structure of epitaxial bismuth telluride topological insulator films correlated to anomalies in charge carrier density\\
  SUPPLEMENTARY MATERIAL}

\author{Stefan Kycia}
%\email[]{skycia@uoguelph.ca}
\affiliation{Department of Physics, University of Guelph, Guelph, Ontario N1G\,1W2, Canada}
\author{S\'ergio L. Morelh\~ao}
%\email[corresponding author:]{morelhao@if.usp.br}
\affiliation{Department of Physics, University of Guelph, Guelph, Ontario N1G\,1W2, Canada}
\address{Institute of Physics, University of S\~ao Paulo, S\~ao Paulo 05508-090, Brazil}
\author{Samuel Netzke}
%\email[]{snetzke@uoguelph.ca}
\affiliation{Department of Physics, University of Guelph, Guelph, Ontario N1G\,1W2, Canada}
\author{Celso I. Fornari}
%\email[]{celsoifornari@gmail.com}
\affiliation{National Institute for Space Research, S\~ao Jos\'e dos Campos, S\~ao Paulo 12227-010, Brazil}
\author{Paulo H. O. Rappl}
%\email[]{rappl@las.inpe.br}
\affiliation{National Institute for Space Research, S\~ao Jos\'e dos Campos, S\~ao Paulo 12227-010, Brazil}
\author{Eduardo Abramof}
%\email[]{eduardo.abramof@inpe.br}
\affiliation{National Institute for Space Research, S\~ao Jos\'e dos Campos, S\~ao Paulo 12227-010, Brazil}

\maketitle
\makeatletter
\renewcommand{\thefigure}{S\@arabic\c@figure}
\renewcommand{\thetable}{I\@Roman\c@figure}
\makeatother

\section{Determination of diffracting domain sizes in thin films}

In small crystals (nanoscale length dimensions), to calculate the intensity distribution around reciprocal lattice nodes and the corresponding line profile of diffraction peaks in rocking curve measurements, each diffraction vector can be conveniently written as $\vec{Q}_{\rm hkl}=(2\pi/d_{\rm hkl})\hat{\vec{e}}_3$ in an arbitrary orthonormal frame of unit vectors $\hat{\vec{e}}_1$, $\hat{\vec{e}}_2$, and $\hat{\vec{e}}_3$. Wavevectors of the incident and diffracted x-rays of wavelength $\lambda$ written as $\vec{k}=(2\pi/\lambda)[\cos\theta\,\hat{\vec{e}}_1 - \sin\theta\,\hat{\vec{e}}_3]$ and $\vec{k}^\prime=(2\pi/\lambda)[\cos\theta^\prime\,\hat{\vec{e}}_1+ \sin\theta^\prime\sin\varphi^\prime\,\hat{\vec{e}}_2+ \sin\theta^\prime\cos\varphi^\prime\,\hat{\vec{e}}_3]$, respectively. Reciprocal vectors ending on the surface of the Ewald sphere are then given by $\vec{Q}=\vec{k}^\prime-\vec{k}$, whose distance $\Delta\vec{Q}=\vec{Q}-\vec{Q}_{\rm hkl}$ from the hkl reciprocal lattice node has the following components
\begin{eqnarray}
\nonumber \Delta Q_1 &=& \Delta\vec{Q}\cdot\hat{\vec{e}}_1 = (2\pi/\lambda)[\cos\theta^\prime - \cos\theta] \\
\nonumber  \Delta Q_2 &=& \Delta\vec{Q}\cdot\hat{\vec{e}}_2 = (2\pi/\lambda)\sin\theta^\prime \sin\varphi^\prime \\
\nonumber  \Delta Q_3 &=& \Delta\vec{Q}\cdot\hat{\vec{e}}_3 = (2\pi/\lambda)[\sin\theta^\prime + \sin\theta]-Q_{\rm hkl}\,.
\end{eqnarray}

By taking $W(\Delta\vec{Q})$ as the Fourier transform of the shape of the crystal, the line profile of the diffraction peak can be calculated as \cite{sm16}
\begin{equation}\label{eq:Itheta}
    I(\theta)=\iint|W(\Delta\vec{Q})|^2\sin\theta^\prime{\rm d}\theta^\prime{\rm d}\varphi^\prime\,.
\end{equation}
Peak maximum occurs at $\theta=\theta^\prime=\arcsin(\lambda/2d_{\rm hkl})$ and $\varphi^\prime=0$ where $d_{\rm hkl}$ is the atomic interplanar distance of Bragg planes.

For a rectangular crystal with edge vectors $L_x\,\hat{\vec{x}}$, $L_y\,\hat{\vec{y}}$, and $L_z\,\hat{\vec{z}}$, and volume $V=L_xL_yL_z$, its Fourier transform is given by the product of three sinc functions,
\begin{equation}\label{eq:WQ1}
W(\Delta\vec{Q})=V\prod_{\alpha=x,y,z} \frac{\sin(\Delta Q_\alpha L_\alpha/2)}{\Delta Q_\alpha L_\alpha/2}\,.
\end{equation}
The line profile also depends on the orientation matrix between diffraction and edge vectors. For rocking curves of asymmetric reflections in non-coplanar diffraction geometry, we choose $L_x$ to be the edge lying on the incidence plane and perpendicular to vector $\vec{Q}$, $L_y$ as the edge in the plane of the film along the projection $\vec{Q}_y$ of the diffraction vector, and $L_z$ as the edge along the growth direction. In terms of the angle $\gamma_Q$ between vector $\vec{Q}$ and the growth direction, the orientation matrix is such that $\hat{\vec{x}}=\hat{\vec{e}}_1$, $\hat{\vec{y}}=\cos\gamma_Q\,\hat{\vec{e}}_2 - \sin\gamma_Q\,\hat{\vec{e}}_3$, and $\hat{\vec{z}}=\sin\gamma_Q\,\hat{\vec{e}}_2 + \cos\gamma_Q\,\hat{\vec{e}}_3$, resulting in
\begin{eqnarray}
\nonumber \Delta Q_x &=& \Delta Q_1 \\
\nonumber  \Delta Q_y &=& \Delta Q_2 \cos\gamma_Q - \Delta Q_3 \sin\gamma_Q \\
\nonumber  \Delta Q_z &=& \Delta Q_2 \sin\gamma_Q + \Delta Q_3 \cos\gamma_Q\,.
\end{eqnarray}

\begin{figure}[t]
  \includegraphics[width=5in]{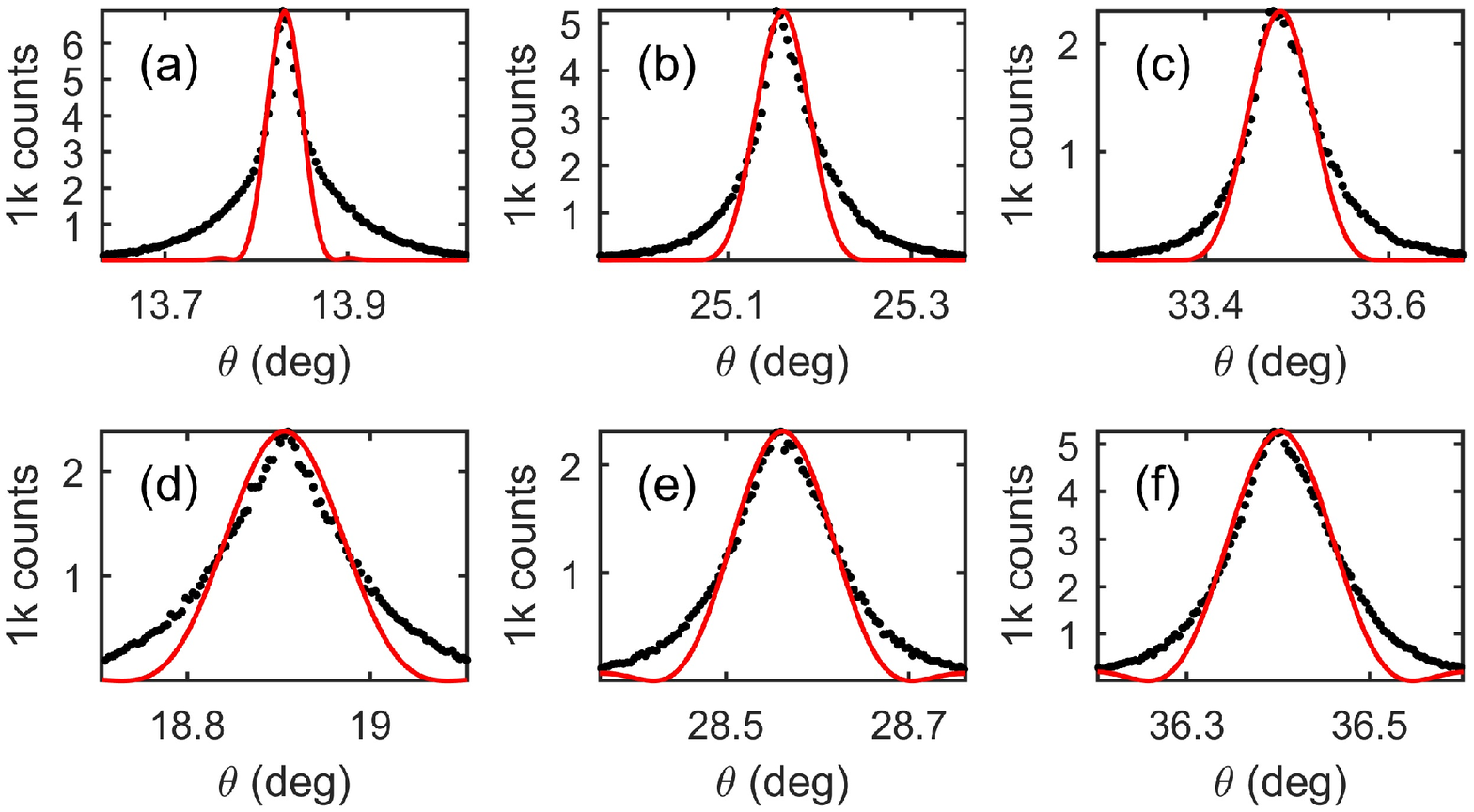}\\
  \caption{Rocking curves of asymmetrical reflections in Bi$_2$Te$_3$ film on BaF$_2$ (111): (a) $\bar{1}0.5$, (b) $0\bar{2}.5$, (c) $\bar{2}\bar{1}.5$, (d) $0\bar{1}.10$, (e) $\bar{2}0.10$, and (f) $1\bar{3}.10$. In-plane, $L_y$, and out-of-plane, $L_z$, mean dimensions of crystal domains in the film are determined by fitting peak widths at half maximum with the line profile function (solid-red line) in Eq.~(\ref{eq:Itheta}). (a-c) $L_y=47.5$\,nm and $L_z=157.5$\,nm. (d-f) $L_y=42.0$\,nm and $L_z=16.3$\,nm. $L_x\simeq5\,\mu$m in all cases.}\label{fig:rcexpsim}
\end{figure}

A simulated annealing (SA) algorithm was used to adjust $L_\alpha$ in Eq.~(\ref{eq:WQ1}) by minimizing $E^2=\sum_{j=1,2,3}(W_e-W_s)_j^2$ where $W_e$ and $W_s$ stand for experimental and calculated peak widths at half maximum, respectively. $W_s$ is obtained from the line profile function $I(\theta)$, Eq.~(\ref{eq:Itheta}). Subscript $j$ runs over either subsets of reflections, i.e. \{$\bar{1}0.5$, $0\bar{2}.5$, $\bar{2}\bar{1}.5$\} or \{$0\bar{1}.10$, $\bar{2}0.10$, $1\bar{3}.10$\}. The results for all samples are presented in Table~\ref{tab:meanL} where the uncertainties were estimated from the error bars in $W_e$ values, $(\Delta L/L)^2 = \sum_{j=1,2,3}(\Delta W_e/W_e)_j^2 $. Examples of experimental rocking curves and calculated profiles using Eq.~(\ref{eq:Itheta}) are shown in Fig.~\ref{fig:rcexpsim} for one sample.

\begin{table}
\caption{Mean-dimensions of crystal domains in Bi$_2$Te$_3$ films on BaF$_2$ (111). Dimension $L_y$ in the plane of the film (along the in-plane projection $\vec{Q}_y$ of the diffraction vector) and dimension $L_z$ along film thickness. Dimension values were determined by using the line profile function in Eq.~(\ref{eq:Itheta}) and a SA algorithm to fit the diffraction peak widths either of hk.5 or hk.10 reflections.}\label{tab:meanL}
\begin{tabular}{cccccc}
  \hline\hline
   & \multicolumn{2}{c}{$L_y$ (nm)} & & \multicolumn{2}{c}{$L_z$ (nm)} \\
   \cline{2-3}\cline{5-6}
Sample & hk.5 & hk.10 & & hk.5 & hk.10 \\
  \hline
S15p &  55.6(4.8)  &  40.4(1.3) & & 114.3(9.8) & 10.5(0.3) \\
S17n & 149.2(9.6)  & 111.5(2.1) & & 136.6(8.8) & 72.5(1.4) \\
S19n & 166.5(13.7) & 163.9(3.0) & &  70.5(5.8) & 54.6(1.0) \\
S27p &  78.8(5.9)  &  85.4(5.2) & & 131.3(9.8) & 51.2(3.1) \\
  \hline
  \hline
\end{tabular}
\end{table}

\section{Lateral lattice mismatch}

Measurement of hybrid reflection pairs is the most accurate method to determine lateral lattice mismatch $\Delta a/a$ in epitaxic films \cite{jzd16,sm18}. In a pair where the hybrid peaks arise from simple sequences of only two reflections, first reflection in the film and the second in the substrate lattice or vice-versa, i.e.,  $\vec{Q}^*=\vec{Q}_f + \vec{Q}_s$ or $\vec{Q}^*=\vec{Q}_s + \vec{Q}_f$ where $\vec{Q}_{f,s}$ stand for diffraction vectors in the film or substrate lattice, the split of hybrid pairs as function of the rocking curve angle $\theta$ is proportional to $\Delta a/a$ as given by \cite{sm18} \begin{equation}\label{eq:thincxphi}
    \Delta\theta \simeq -2\frac{\vec{Q}_{f,\|}\cdot\hat{\vec{k}_{\|}}}{Q^*}\frac{\Delta a}{a}\,.
\end{equation}
$\vec{Q}_{f,\|}$ is in-plane projection of the film diffraction vector and $\hat{\vec{k}_{\|}}$ is the in-plane direction of the incident wavevector. For the measured pair in this work, hybrids $\bar{2}2.\bar{10}_f+044_s$ (peak 1) and $404_s+0\bar{2}.\bar{10}_f$ (peak 4) in Fig. 2(c) (main text),   $\vec{Q}_{f,\|}\cdot\hat{\vec{k}_{\|}}/Q^*=1.0176$, leading to the values presented in Table I (main text).

%\bibliography{lateraldisorderBi2Te3suppinfo}